\documentclass[12pt]{iopart}
\pdfoutput=1

\expandafter\let\csname equation*\endcsname\relax
\expandafter\let\csname endequation*\endcsname\relax
\usepackage[english]{babel}
\usepackage[ansinew]{inputenc} 
\usepackage[T1]{fontenc}	
\usepackage[]{graphicx} 
\usepackage{amsmath} 
\usepackage{dsfont} 
\usepackage{pgf} 
\usepackage[colorlinks=true,allcolors=black]{hyperref}

\newcommand{\bra}[1]{\ensuremath{\left\langle #1\right|}}
\newcommand{\ket}[1]{\ensuremath{\left|#1\right\rangle}}

\begin{document}
\graphicspath{{pictures/}}
\title[Quantum metrology in spatially correlated noise]{Quantum metrology subject to spatially correlated Markovian noise: \\Restoring the Heisenberg limit}

\author{Jan Jeske and Jared H. Cole}
\address{Chemical and Quantum Physics, School of Applied Sciences, RMIT University, Melbourne, 3001, Australia}
\author{Susana F. Huelga}
\address{Institut f\"ur Theoretische Physik, Albert-Einstein-Allee 11, Universit\"at Ulm, D-89069 Ulm, Germany}

\begin{abstract}
Environmental noise can hinder the metrological capabilities of entangled states. While the use of entanglement allows for Heisenberg-limited resolution, the largest permitted by quantum mechanics, deviations from strictly unitary dynamics quickly restore the standard scaling dictated by the central limit theorem. Product and maximally entangled states become asymptotically equivalent when the noisy evolution is both local and strictly Markovian. However, temporal correlations in the noise have been shown to lift this equivalence while fully (spatially) correlated noise allows for the identification of decoherence-free subspaces. Here we analyze precision limits in the presence of noise with finite correlation length and show that there exist robust entangled state preparations which display persistent Heisenberg scaling despite the environmental decoherence, even for small correlation length. Our results emphasize the relevance of noise correlations in the study of quantum advantage and could be relevant beyond metrological applications.
\end{abstract}

\maketitle

\section{Introduction}
How small can experimental error bars become? In classical and standard quantum experiments increasing the number of measurements $n$ decreases the uncertainty as $1/\sqrt{n}$, the standard quantum limit (SQL). The use of entangled states in quantum metrology shows that setups exist with an improved scaling $1/n$, the Heisenberg limit. This defines {\em quantum advantage}, that is, exploiting a quantitative (quantum) resource to outperform classical strategies. While quantum advantage has been proven impossible for uncorrelated Markovian noise, we show in this article, that Heisenberg scaling can be restored in the presence of spatially correlated Markovian noise.

In a typical metrological set up, the estimation of a given parameter (like an external field) is mapped out into the estimation of a transition frequency. This is done by measuring the phase relation accumulated between the two components in a superposition state. In a purely unitary evolution, the relative phase of a Greenberger-Horne-Zeilinger (GHZ) state of the form $(\ket{0}^{\otimes n} + \ket{1}^{\otimes n})/\sqrt{2}$ advances $n$ times faster than that of a single qubit and parity measurements allow to saturate the Heisenberg limit \cite{Wineland1992,Wineland1994}. That is, the associated measurement uncertainty decreases as $1/n$, which provides a $1/\sqrt{n}$ improvement over the SQL obtained by performing $n$ independent queries on uncorrelated particles. Signal-to-noise ratios (SNR) overcoming the spectroscopic resolution achievable in ideal experiments using single qubits have been demonstrated using three entangled ions \cite{leibfried2004}. When the dimension of the probe state grows, decoherence effects will no longer be negligible and the question arises of whether Heisenberg scaling can still be attainable under a non unitary (noisy) evolution.
In the case of the dynamics being strictly Markovian, and provided the noise is local, both pure dephasing and dissipative losses restore the standard scaling dictated by the central limit theorem even in the limit
of arbitrarily small noise levels \cite{Huelga1997}. Rigorous bounds showing standard scaling under this type of noise have been recently put forward \cite{Escher2011,Demkowicz2012}.

These noise models though make two important assumptions, namely, that the noise stems from a Markovian bath and that this noise acts locally on each subsystem. Relaxing the assumption of Markovianity has been shown to result in a new fundamental limit \cite{Chin2012, Matsuzaki2011} which lifts the previous metrological equivalence of maximally entangled and product states under time-correlated noise and predicts a novel scaling of the form $\sim 1/n^{3/4}$.  In this paper we relax the assumption of noise locality and consider a general model for bath correlation \emph{length} \cite{Jeske2013formalism} to show the persistence of Heisenberg scaling under correlated noise of finite length.

\section{Overview of precision scalings}
We will consider a standard set up for performing Ramsey-type interferometry \cite{Wineland1994} where the input (probe) systems are entangled states of $n$ qubits of the form $\ket{\Psi(t)}=(\ket{\alpha} +\exp(i \omega t) \ket{\beta})/\sqrt{2}$, where $\ket{\alpha}$ and $\ket{\beta}$ are $n$-qubit product states. In the presence of Markovian dephasing noise, the dynamics of such states is fully characterized by the time evolution the off-diagonal density matrix element $\rho_{\alpha \beta}$. This matrix element, often referred to as the system's coherence, typically shows a time-evolution of the form \cite{Braun2006, Hein2005}:
\begin{align}
\rho_{\alpha \beta}(t)=\rho_{\alpha \beta}(0) \exp[-i n \omega_0 t - \Gamma(n,\xi) t] \label{off-diagonal time evolution}
\end{align}
where $\omega_0$ is the qubit transition frequency and $\Gamma(n,\xi)$ is the dephasing rate of the entangled state, dependent on the number $n$ of considered qubits and the spatial correlation length $\xi$ of the environmental noise. The functional form of $\Gamma(n,\xi)$, which scales with $n$ differently for different types of noise, will determine the achievable precision of the metrological scheme, as we will show later.
In the special case of completely uncorrelated dephasing, $\Gamma(n,0)=n \gamma $ and eq.~\ref{off-diagonal time evolution} coincides with the results in \cite{Huelga1997} for the decoherence of GHZ states.

In a Ramsey interferometer, the atomic transition frequency $\omega_0$ can be obtained from the above time evolution by means of interrogating the system with simple projective measurements. The resulting measurement uncertainty can be calculated in terms of the classical Fisher information $F(\omega_0)$ \cite{Braunstein1994}:
\begin{align}
\Delta \omega_0=\sqrt{\frac{1}{N F(\omega_0)}} \qquad
F(\omega_0)=\sum_{j=1,2} \frac{1}{p_j} \left( \frac{\partial p_j}{\partial \omega_0} \right)  ^2
\end{align}
where $N$ is the number of repetitions of the experiment and $p_j$ the probabilities of the two outcomes of the experiment.
The probabilities corresponding to eq.~\ref{off-diagonal time evolution} are $p_1=[1+\cos(n \omega_0 t) \exp(-\Gamma(n,\xi) t) ]/2$ and $p_2=1-p_1$. They are obtained as the probabilities of the +1 and -1 eigenvalues of the measurement operator $P=\ket{\alpha}\bra{\beta} + h.c.$. The measurement uncertainty from these probabilities is still dependent on the interrogation time $t$.  Considering as fixed resources the total number of particles $n$ and the total duration of the experiment $T$, the duration of the optimal interrogation time $t$, and with it the number of repetitions $N=T/t$ for a given noise source has to be determined \cite{Huelga1997}. Note that the time required for initialisation and measurement are neglected in this discussion; the persistence of quantum advantage when these are taken into account is discussed in \cite{Cirac1999}.  Minimizing the measurement uncertainty with $t$ first yields a closely spaced set of times $t=m \pi / (2 n \omega_0)$, with $m$ odd, which minimizes the uncertainty in each oscillation. Out of these we choose the one which is approximately at the minimum of the envelope uncertainty, at $t=1/[2 \Gamma(n,\xi)]$. This yields the time-optimized frequency uncertainty of the measured frequency
\begin{align}
\Delta \omega_0 = \sqrt{\frac{2 e \Gamma(n, \xi)}{n^2 T}}.\label{frequency uncertainty}
\end{align}
This value is bounded from below by the quantum Cramer Rao bound \cite{Braunstein1994}. In our case, we are concerned with the persistence of Heisenberg scaling in the presence of correlated noise and therefore do not consider the issue of optimizing the final measurement procedure to obtain the ultimate precision. That is, we do not consider the possible reduction of merely numerical factors within the same functional dependence on $n$.

For spatially uncorrelated Markovian decoherence, the dephasing rate of GHZ states scales as $\Gamma(n,\xi) \rightarrow \Gamma_{\rm uc} =n \gamma$, where $\gamma$ is the single qubit dephasing rate. This yields an SQL resolution $\sim 1/\sqrt{n}$ (eq.~\ref{frequency uncertainty}). The persistence of the standard scaling under uncorrelated Markovian decoherence is even valid for optimized (entangled) initial states and generalized measurements, with the optimal achievable resolution
$
\Delta \omega_0^{opt} = \sqrt{2 \gamma / (n T)}
$ \cite{Huelga1997,Escher2011}. To go beyond SQL scaling the dephasing rate would need to scale with a power less than 1.  Furthermore, for Heisenberg scaling the rate would need to be independent of $n$.

While the derivation above relies on an {\em independent noise model} \cite{Palma1996, Reina2002}, recent experiments with trapped ions have proven to be dominated by spatially correlated dephasing \cite{Chwalla2007, Monz2011}. Particularly, measurements of the dephasing rate of GHZ states have shown a clear $n^2$ dependence \cite{Monz2011}, a form of ``superdecoherence'', which can only be explained by strongly correlated noise. When the noise on all qubits is perfectly correlated, it is possible to identify suitable decoherence-free subspaces as discussed theoretically in \cite{Dorner2012}; we generalise the concept of correlated noise and introduce for the first time a correlation length into the noise model. In correlated noise certain subspaces show a decoherence rate which reduces with increasing correlation length and become decoherence-free for fully correlated environments. This allows for the accurate determination of frequencies with a decoherence-reduced encoded state whose existence depends on the finite correlation length of the bath. In \cite{Roos2006} a decoherence-reduced subspace has been demonstrated experimentally by using an entangled state of two ions for the determination of the quadrupole moment of $^{40}$Ca$^{+}$, a quantity of relevance for the calibration of optical frequency standards \cite{itano2006}. We address particularly in this paper what the situation would be when considering larger qubit arrays so that the noise exhibits a correlation length that is smaller than the total length of the system. Using a formalism to consider realistic partially correlated noise, where the correlations decay over a certain correlation length $\xi$ \cite{Jeske2013formalism}, we will show that, even for small $\xi$, Heisenberg scaling prevails when using certain types of entangled states for the estimation of small frequency shifts, as those involved for instance in the precise estimation of an atomic quadrupole moment. We use the measurement of the quadrupole moment as an illustrative example but the procedure is more general and could also be applicable to sensing external fields.

\section{The model}
\begin{figure}
\centering
\includegraphics[scale=1]{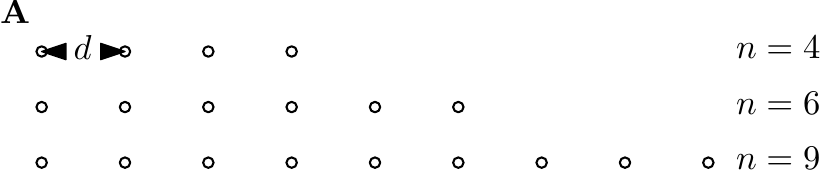}\\
\hspace{0.28cm} \includegraphics[scale=1]{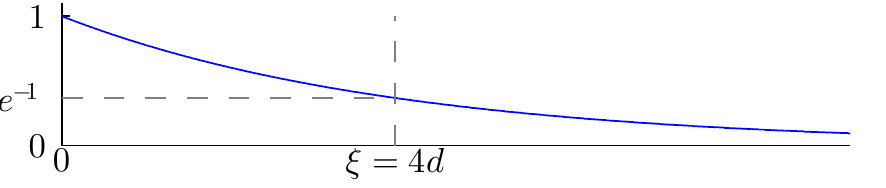} \vspace{0.7cm}\\
\hspace{.8cm}\includegraphics[scale=1]{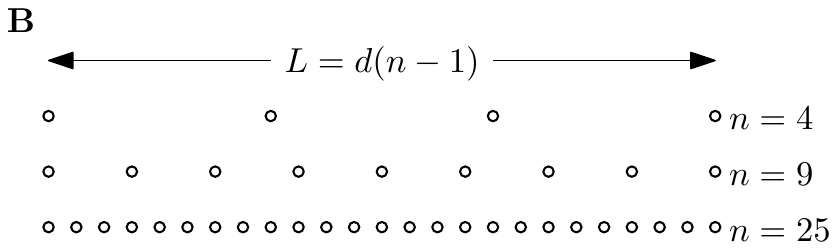}\\
\includegraphics[scale=1]{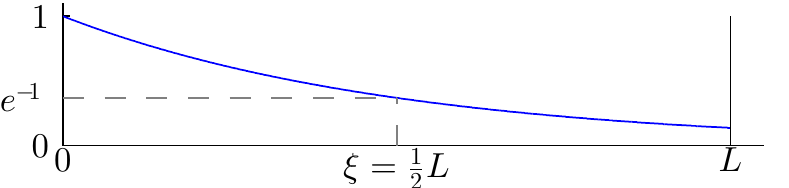}
\caption{Three arrays with increasing numbers $n$ of ions with two ways of scaling them relative to the spatial correlation function $C(0,x)=\exp(-x/\xi)$: \textbf{A} The distance $d$ between ions is fixed and the array becomes longer with increasing $n$. \textbf{B} The array length $L$ is fixed and ions become more dense with increasing number $n$. }
\label{fig Array with L and d}
\end{figure}

We consider a system of $n$ hydrogen-like ions with a Zeeman splitting term of the sublevels $J_z \ket{m}=\hbar m \ket{m}$ of the total angular momentum $J_z$ and a small correction term due to the interaction of the atomic electric quadrupole moment with the external electric field gradient. These correction terms are quadratic in $J_z$ \cite{Roos2006, Itano2000} and we will focus on entangled states with a relative frequency due to these correction terms.
\begin{align}
H_s=\beta \sum_{j=1}^n J_z^{(j)} + \alpha \sum_{j=1}^n (J_z^{(j)})^2
\label{Hamiltonian}
\end{align}
Laser frequency noise and magnetic field noise make dephasing by far the strongest decoherence source, effectively coupling each ion to a fluctuation via $J_z$,
$
H_{int}= v \sum_j J_z^{(j)} B_j.
$
The coupling strength $v$ defines the total decoherence strength by the coefficient $\gamma_0=v^2$ in all dephasing rates, so for simplicity we set $v=1$. The bath operators' $B_j$ spatial and temporal correlations are determined by the function $C(\omega, x_j-x_k)= \int_{-\infty}^\infty e^{i \omega \tau} \langle \tilde B_j(\tau,x_j) \tilde B_k(0,x_k) \rangle$, where we assume the ions are spatially arranged in a linear array (figure \ref{fig Array with L and d}). We employ Bloch-Redfield equations, which simplify further because $H_s$ and $H_{int}$ commute. We assume homogeneous, decaying spatial correlations $C(0,xd)=\exp(-|xd|/\xi)$ with the correlation length $\xi$, the distance $d$ between ions in the one-dimensional array and $x \in \mathds{N}$. The mathematical form of the Bloch-Redfield equations does not guarantee completely positive time evolution, which is a physically necessary condition. However a consistent model of the spatial and temporal correlations will guarantee that complete positivity is not broken. We showed explicitly in reference \cite{Jeske2013formalism}, that the chosen exponential spatial correlations can always be mapped to Lindblad form and therefore obey complete positivity. We arrive at the master equation for the system density matrix $\rho$ as \cite{Jeske2013formalism}:
\begin{align}
\dot \rho = \frac{i}{\hbar} [\rho,H_s] + \frac{1}{\hbar^2} \frac{1}{2} \sum_{j,k} \exp\left( -\frac{|x_j-x_k| d}{\xi} \right) \left( 2J_z^{(j)} \rho J_z^{(k)} - J_z^{(j)} J_z^{(k)} \rho - \rho J_z^{(j)} J_z^{(k)} \right) \label{master equation}
\end{align}

As initial states we consider entangled states of the form $(\ket{m_1, m_2,\dots,m_n} + \ket{\tilde m_{1},\tilde m_2, \dots, \tilde m_{n}}) / \sqrt{2}$ where the magnetic quantum numbers $m_j$ of the operators $J_z^{(j)}$ satisfy $\sum_{j=1}^n m_j = \sum_{j=1}^{n} \tilde m_j$. The two parts of the superposition are Zeeman-shifted by the same amount but their quadrupole moment can be different $\sum_{j=1}^n m_j^2 \neq \sum_{j=1}^{n} \tilde m_j^2$. In other words the two components of the entangled states have a relative frequency only due to the second term in the Hamiltonian eq.~\ref{Hamiltonian}. This is the defining property for the state of interest and it can be implemented using different numbers of sublevels of $J_z$. However for simplicity here we restrict ourselves to three sublevels: $m , \tilde m \in \{\epsilon_+,\epsilon_0,\epsilon_-\}$ where $\epsilon_+ = \epsilon_0 + \epsilon_\Delta$ and $\epsilon_- = \epsilon_0 - \epsilon_\Delta$. For example in ref.~\cite{Roos2006} the levels $m , \tilde m  \in \{3/2, -1/2,-5/2\}$ in $^{40}$Ca$^+$ ions were used. We choose for one part of the initial entangled state all ions to be in $\epsilon_0$ and in the other part half of the ions in $\epsilon_+$ and half in $\epsilon_-$, i.e.~$(\ket{\epsilon_+, \epsilon_+, \dots, \epsilon_-,\epsilon_-,\dots} + \ket{\epsilon_0,\epsilon_0,\epsilon_0,\dots})/\sqrt{2}$. This simplifies $J_z=\rm diag(\epsilon_+, \epsilon_0, \epsilon_-)$ and the coherent evolution is given by $\ket{\Psi(t)}= (\ket{\epsilon_+, \epsilon_+, \dots, \epsilon_-,\epsilon_-,\dots} + \exp(i n \omega_0 t)\ket{\epsilon_0,\epsilon_0,\epsilon_0,\dots})\sqrt{2}$ where the relative frequency is given by the quadrupole splitting $\omega_0=\alpha \epsilon_\Delta^2$. This frequency is measured with a parity measurement and we will regard the uncertainty scaling with $n$ of this transition frequency. The probabilities of even and odd parity measurement can be calculated by the probabilities of the $+1$ and $-1$ eigenvalues of the operator $P=\ket{\epsilon_+, \epsilon_+, \dots, \epsilon_-,\epsilon_-,\dots}\bra{\epsilon_0,\epsilon_0,\epsilon_0,\dots}+h.c.$. For experimental implementations of parity measurements see refs.~\cite{Wineland1992, Roos2006}

\section{Perfect spatial correlations}
Uncorrelated Markovian decoherence ($\xi\rightarrow0$) always restores the standard quantum limit \cite{Huelga1997}, whereas for the chosen states correlated decoherence leads to Heisenberg scaling:
For perfect correlations, i.e.~infinite correlation length $\xi \rightarrow \infty$ the spatial correlation function in the master equation \eqref{master equation} becomes $\exp(x/\xi) \rightarrow 1$.
Taking the sums into each term we define a new Hermitian operator $S=\sum_j J_z^{(j)}$ with $S^\dagger=S$. We therefore find the equations to be of the simple Lindblad form:
\begin{align}
\dot \rho = \frac{i}{\hbar} [\rho,H_s] + \frac{1}{\hbar^2} \left( S \rho S - \frac{1}{2} \{S^2,\rho\} \right)
\end{align}
The Hermitian operator $S$ yields the same value for all states with the same number of excitations. Particularly for the two constituents of our initial state we find:
\begin{align}
S \ket{\epsilon_+,\epsilon_+,\dots,\epsilon_-,\epsilon_-,\dots} = S \ket{ \epsilon_0,\epsilon_0,\epsilon_0,\dots} \label{S acts equally}
\end{align}
In other words for their subspace $S \propto \mathds{1}$ and the master equation becomes:
\begin{align}
\dot \rho &= \frac{i}{\hbar} [\rho,H_s] + \frac{1}{\hbar^2} \left( \mathds{1} \rho \mathds{1} - \frac{1}{2} \{\mathds{1}^2,\rho\} \right) 
=\frac{i}{\hbar} [\rho,H_s] + 0
\end{align}
This means our entangled initial state is in a decoherence-free subspace (for a perfectly correlated bath). Its time evolution is given by:
\begin{align}
\ket{\Psi(t)} &= \ket{\epsilon_+,\epsilon_+,\dots,\epsilon_-,\epsilon_-,\dots} + e^{i n \omega_0 t} \ket{\epsilon_0,\epsilon_0,\epsilon_0,\dots}
\end{align}
Since eq.~\ref{S acts equally} holds for all pairs of states with equal excitation number we find that the order of ions is irrelevant for perfect correlations \cite{Dorner2012}.

\section{Partial spatial correlations}
We will now discuss the persistence of Heisenberg scaling under partially correlated noise and assume a finite but non-vanishing correlation length $\xi>d$. The time-evolution of the coherence of interest $\rho_{\pm,0}$ takes the form of eq.~\ref{off-diagonal time evolution} (\ref{app partial correlations dephasing rate}):
\begin{align}
\rho_{\pm,0}(t)=\rho_{\pm,0}(0)\exp(-in\omega_0 t - \Gamma(n,\xi) t)
\end{align}
where the dephasing rate $\Gamma(n,\xi)$ depends on the number of ions $n$ involved in the superposition and the correlation length $\xi$. In stark contrast to uncorrelated decoherence, it also depends on the \emph{order} of the ions in the initial state. This can be understood by regarding the decoherent part of eq.~\ref{master equation}:
\begin{align}
\sum_{j,k} \exp\left( -\frac{|x_j-x_k| d}{\xi} \right) \left( 2J_z^{(j)} \rho J_z^{(k)} - J_z^{(j)} J_z^{(k)} \rho - \rho J_z^{(j)} J_z^{(k)} \right) \label{sum in master equation}
\end{align}
All pairs $j,k$ of ions contribute terms to $\Gamma(n,\xi)$. Some terms increase the dephasing rate, others decrease it depending on the sign of the second factor. Pairs of ions which are in the same state (i.e.~both $\epsilon_+$ or both $\epsilon_-$) increase the dephasing rate (cf.~case 2 in \ref{app partial correlations dephasing rate}), while pairs in opposite states compensate by reducing the dephasing rate (cf.~case 1 in \ref{app partial correlations dephasing rate}). Autocorrelations only increase the dephasing rate (compare case 3). The order of the ions changes the first factor in eq.~\ref{sum in master equation}. Since the first factor decreases with distance, a state is more robust the closer pairs with opposite states are in space. We therefore reorder the ions in the initial state to the new order $(\ket{\epsilon_+,\epsilon_-,\epsilon_+,\epsilon_-,\dots}+\ket{\epsilon_0,\epsilon_0,\epsilon_0,\dots})/\sqrt{2}$. The dephasing rate corresponding to this state is given by (see appendix~\ref{app partial correlations dephasing rate}):
\begin{align}
\Gamma(n,\xi)&=\frac{n \epsilon_\Delta^2}{2} - \sum_{x=1}^{n/2} (n-2x+1) C(0,(2x-1)d)\epsilon_\Delta^2 
+ \sum_{x=1}^{n/2} (n-2x) C(0,2x d) \epsilon_\Delta^2\label{dephasing rate for general C}
\end{align}
With the assumed correlation function $C(0,xd)=\exp(-|x|d/\xi)$ one finds:
\begin{align}
\Gamma(n,\xi)=\frac{2 e^{\frac{d}{\xi }}-2 e^{\frac{(1-n)d}{\xi }}-n+e^{2d/\xi } n}{2 \left(1+e^{\frac{d}{\xi }}\right)^2} \epsilon_\Delta^2 \label{dephasing rate of n and xi}
\end{align}
To judge whether entangled states give an advantage over the standard quantum limit ($\propto 1/\sqrt{n}$) we need to determine whether the dephasing rate $\Gamma(n,\xi)$ scales faster or slower than $n$ (see eq.~\ref{frequency uncertainty}). For comparison, the dephasing rate for uncorrelated decoherence is $\Gamma_{\rm uc}=n \epsilon_\Delta^2/2$ (see appendix~\ref{app partial correlations dephasing rate} or \cite{Huelga1997}).

The frequency $\omega_0=\alpha\epsilon_\Delta^2$ cannot be measured in a single ion because there are no two states in a single ion with the same contribution from the first term in eq.~\ref{Hamiltonian}; one needs at least an entangled state of two ions to realise it\footnote{We should mention here that in correlated noise environments one can take advantage of the system's superdecoherence which produces this entangled state from a prepared product state \cite{Chwalla2007} at the expense of losing signal contrast. }. To obtain a meaningful comparison for the scaling of the frequency uncertainty this must be taken into account. We therefore define two entangled ions as the minimum entanglement resource for measuring a quadrupole moment.  We then compare the scaling of the $n$--entangled state with a product state of $n/2$ entangled pairs which contribute $n/2$ more measurements to the statistics. For this minimal entangled array of ion pairs, we find an uncertainty that scales with the SQL as we increase the number $n/2$ of pairs, $\Delta \omega_{0,\rm p}= \sqrt{ e \Gamma(2,\xi)/(n T) }$.

In contrast to uncorrelated decoherence, there is no unique way of increasing the number of entangled ions for a given noise correlation length.  When considering the scaling of the dephasing rate with increasing numbers of ions, one can either keep the \emph{length} $L$ of the ion array fixed, or the \emph{density} of ions fixed, as can be achieved in segmented traps \cite{Kielpinski2002}. We now analyze the achievable spectroscopic resolution in both cases (figure \ref{fig Array with L and d}).

\begin{figure}
\centering
\includegraphics[scale=0.86]{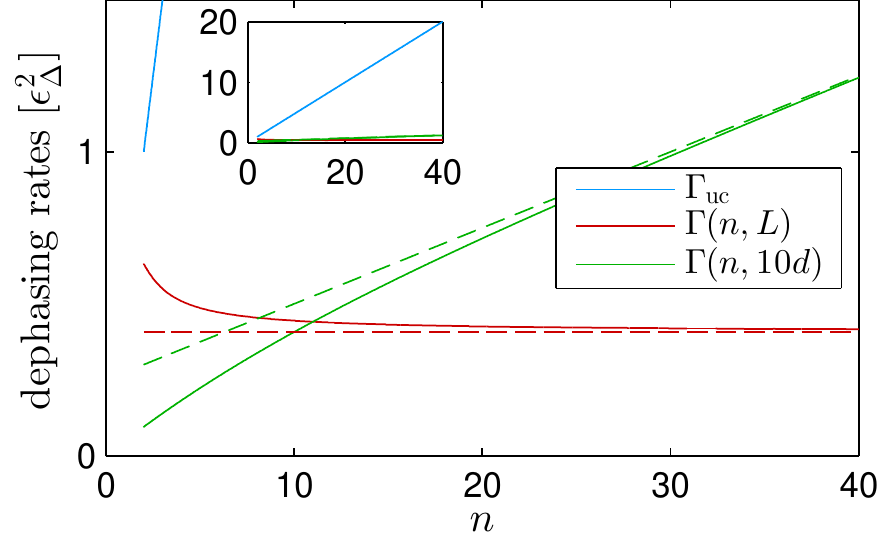}
\includegraphics[scale=0.87]{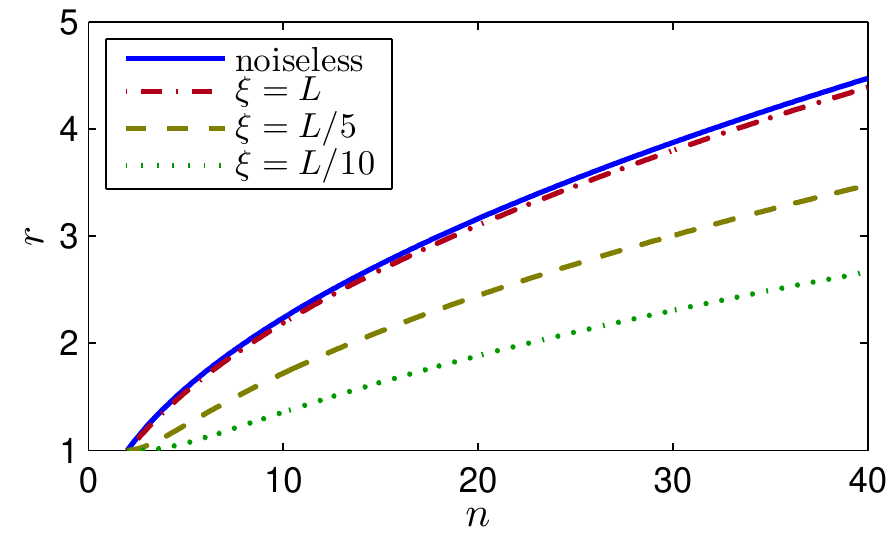}
\caption{\textbf{Left: }Dephasing rates in units of $\epsilon_\Delta^2$ and their scaling with $n$ for uncorrelated decoherence (linear, steep), $\xi=L$ (constant) and $\xi=10d$ (linear, small gradient). Dashed lines are the approximations for large $n$. This plot shows that even partially correlated decoherence is strongly advantageous for the chosen initial states. \textbf{Right: }Relative frequency resolution $r$ of a partially correlated environment with $\xi=L/10$ (dot-dashed), $\xi=L/5$ (dashed), $\xi=L$ (dotted) and a noiseless environment (solid). An $n$--entangled state scales better by a factor of $\sqrt{n}$ than a pair-wise entangled state and approaches noiseless scaling for increasing correlation length $\xi$.}
\label{fig RatesPlot}
\label{fig Heisenberg scaling}
\end{figure}

First we set the correlation length to a fixed number of ions $\xi=c d$ (figure \ref{fig Array with L and d}\textbf{A}) which means that the array gets longer relative to the correlation length as we increase $n$ (fixed ion density). This will ultimately restore the SQL when $n d \gg \xi$ (figure \ref{fig RatesPlot}). The gradient of the dephasing rate in this case (green line in figure \ref{fig RatesPlot}) can be approximated as $\epsilon_\Delta^2 d/(4\xi)$ for $\xi>d$. This gradient is smaller than for uncorrelated decoherence $\Gamma_{\rm uc}$ because the finite correlation length reduces the dephasing-rate contribution from each ion slightly. So even though the scaling follows the standard quantum limit, one finds a better coefficient than for uncorrelated decoherence.

Alternatively, we can scale the correlation length as a fraction $c$ of the whole array $\xi = c L$ (figure \ref{fig Array with L and d}\textbf{B}), fixing the correlations between the first ion and the last ion in the array to a value $C(0, L) =\exp(-1/c)$. Figure \ref{fig RatesPlot} shows that in this case the dephasing rate quickly approaches the constant $\Gamma(n,\xi)=[-L/\xi+\exp(-L/\xi)-1]\epsilon_\Delta^2/4$, which can be approximated as $\epsilon_\Delta^2 L / (2\xi)$ for long correlation lengths $\xi\gg L$. With this constant rate the corresponding uncertainty (Eq.~\ref{frequency uncertainty}) displays Heisenberg scaling (figure \ref{fig Heisenberg scaling}):
\begin{align}
\Delta \omega_0 \rightarrow \sqrt{\frac{2 e [-L/\xi +\exp(-L/\xi)-1]\epsilon_\Delta^2 }{4 T}} \frac{1}{n} 
\end{align}

We now introduce the relative frequency resolution:
\begin{align}
r= \frac{\Delta \omega_{0,\rm p}}{\Delta \omega_0}  \label{relative frequency uncertainty}
\end{align}
where the full expression for $r$ is given by equations \ref{frequency uncertainty}, \ref{dephasing rate of n and xi} and \ref{relative frequency uncertainty}. We find that with increasing correlation length $\xi$ the uncertainty approaches the noiseless Heisenberg scaling (figure \ref{fig Heisenberg scaling}). Even for partial correlations, which decay on the length scale of the array, the Heisenberg scaling of the uncertainty is robust.

\section{GHZ states subject to correlated noise}
Heisenberg scaling in spatially correlated environments can generally be achieved for entangled states which are superpositions of states with the same number of excitations. A counter--example however are GHZ states. In a noiseless environment a frequency measurement will show Heisenberg scaling of the uncertainty $1/(n\sqrt{Tt})$. For spatially uncorrelated noise their dephasing rate scales with $n$, leading (as for the previously considered initial states) to the SQL for the uncertainty $\sqrt{2 e \gamma/(n T)}$. In spatially correlated noise GHZ states are even more fragile, their dephasing rate scales with $n^2$, leading to an uncertainty $\sqrt{2 e \gamma/T}$, which no longer decreases with $n$ at all \cite{Dorner2012}. In spatially correlated noise environments GHZ states are therefore strongly disadvantageous.

From our master equation (Eq.~\ref{master equation}) it follows generally that for $\xi\rightarrow\infty$ the dephasing rate between two states with a difference of $n_e$ excitations is proportional to $n_e^2$. States with the same number of excitations have $n_e=0$ and form a decoherence-free subspace, whereas GHZ states have $n_e=n$ and are the most fragile states in spatially correlated environments.

Up to now we have considered perfect spatial correlations, $C(0,x)=1$, and decaying spatial correlations, $C(0,x)=\exp(-x/\xi)$. Both are positive functions for all $x$, and the statements of the last two paragraphs are only valid under this condition. The $n^2$ scaling of the dephasing rates found experimentally for GHZ states indicates that in ion traps these two functional forms are good approximations for the noise correlations. However, it is also physically possible for the spatial correlations to take the homogeneous form $C(0,x)=\cos(k x)$, where points at \emph{specific} distances have noise with negative correlations. For example in reference \cite{Jeske2013formalism} an environmental model of coupled harmonic oscillators is discussed that yields cosine shaped spatial correlations. In such an environment GHZ states can be engineered to be within a decoherence-free subspace by arranging an array of sites such that the array length $L$ matches the oscillation length of the environmental spatial correlations $L=2 \pi/k$ (see appendix~\ref{app oscillating correlations}). Note that uniformly negative correlation functions are impossible due to the requirements of positive autocorrelations and multipartite correlation rules.

\section{Conclusion}
In conclusion, we discussed the impact of spatial noise correlations on quantum metrology and the persistence of Heisenberg scaling in the presence of such noise.  We consider ion traps as one specific example since previous experimental evidence suggests that their environmental noise is spatially correlated with purely positive correlations. We showed that non-zero spatial correlation length fundamentally changes the decoherence of entangled states. In such environments, a {\em topology} dependence emerges so that the order in which the ions are placed in the array changes their decoherence properties. After optimisation in this regard, the entangled states designed to measure the electric quadrupole moment have an approximately constant dephasing rate with increasing number of ions $n$. Precision frequency measurements with these initial states therefore show Heisenberg scaling of the uncertainty $\Delta \omega_0 \propto 1/n$ with the numbers of ions $n$. Besides providing a prescription to achieve Heisenberg-scaled resolution in linear ion traps subject to partially correlated noise, our results illustrate the fundamental role of noise correlations in precision spectroscopy. While local Markovian noise eliminates quantum advantage, this is restored when the noise displays a spatial structure. Heisenberg resolution becomes then attainable by means of suitable state preparation whose decoherence rate decreases inversely with $\xi$ so that the evolution is decoherence-free in the limit of infinite correlation length (global noise).


\ack
We thank A.~Greentree, N.~Vogt and T.~Dubois for helpful discussions,  M.~B.~Plenio for feedback on the manuscript and J.~Home and J.~Hecker-Denschlag for helpful information about noise sources in ion traps. We acknowledge financial support from the European Commission through the STREP project PAPETS.

\appendix
\section{Partial correlations and its dephasing rate for $n$ ions}
\label{app partial correlations dephasing rate}
\begin{figure}[htb]
\centering
\includegraphics{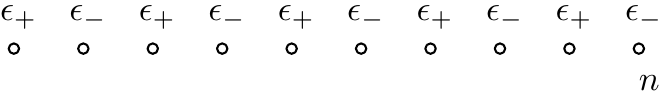}
\caption{A chain of n ions. The operator $J_z^{(j)}$ acting from the left onto our element $\rho_{\pm,0}$ gives $\epsilon_+=\epsilon_0+\epsilon_\Delta$ for $j$ odd and $\epsilon_-=\epsilon_0-\epsilon_\Delta$ for $j$ even.}
\label{fig chain of n ions 2}
\end{figure}

For finite but non-vanishing correlation length $\xi$ we solve the master equation \cite{Jeske2013formalism}
\begin{align}
\dot \rho = \frac{i}{\hbar} [\rho,H_s] + \frac{1}{\hbar^2} \frac{1}{2} \sum_{j,k} C(0,x_j-x_k) \left( 2J_z^{(j)} \rho J_z^{(k)} - J_z^{(j)} J_z^{(k)} \rho - \rho J_z^{(j)} J_z^{(k)} \right) \label{master equation supp mat}
\end{align}
and derive the dephasing rate explicitly for an exponentially decaying spatial correlation function $C(0,x_j-x_k)=\exp\left( -|x_j-x_k|/\xi \right)$. Note that in the appendix we give $\xi$ in units of $d$ and thereby avoid the appearance of $d$ in the enumerator. The density matrix element of interest is:
\begin{align}
\rho_{\pm,0}&=\ket{\epsilon_-,\epsilon_+,\epsilon_-,\epsilon_+,...}\bra{\epsilon_0,\epsilon_0,\epsilon_0,...,\epsilon_0}
\end{align}
The coherent part is easily calculated:
\begin{align}
i (\rho_{\pm,0} H_s - H_s \rho_{\pm,0})= i (\beta n \epsilon_0 + \alpha n \epsilon_0^2 - \beta n \epsilon_0 - \alpha n (\epsilon_0^2 +\epsilon_\Delta^2)\rho_{\pm,0} = -i n \alpha \epsilon_\Delta^2 \rho_{\pm,0} = -i n \omega_0 \rho_{\pm,0}
\end{align}
We find that the relative frequency between the two components of the entangled state scales linearly with $n$ as given in eq.~\ref{off-diagonal time evolution}.

To calculate the decoherent part of eq.~\ref{master equation supp mat} for the element $\rho_{\pm,0}$ we regard how $J_z^{(j)}$ acts from the left and from the right onto element $\rho_{\pm,0}$ (figure \ref{fig chain of n ions 2}):
\begin{align}
J_z^{(j)} \rho_{\pm,0} &= \left\{
\begin{aligned} \epsilon_- \rho_{\pm,0} = (\epsilon_0-\epsilon_\Delta) \rho_{\pm,0} &\quad \text{if $j$ is odd} \\
\epsilon_+ \rho_{\pm,0} = (\epsilon_0+\epsilon_\Delta) \rho_{\pm,0} &\quad \text{if $j$ is even} \end{aligned} \right. \\
\rho_{\pm,0} J_z^{(j)} &= \epsilon_0 \rho_{\pm,0}
\end{align}
The fact that acting the $J_z^{(j)}$ operator on the element reproduces it already shows us, that in eq.~\ref{master equation supp mat} the time derivative of the element only depends on the element itself and can be written in an easily solvable form:
\begin{align}
\dot \rho_{\pm,0} &= -i n \alpha \epsilon_\Delta^2 \rho_{\pm,0} - \Gamma(n,\xi) \rho_{\pm,0}\\
\rho_{\pm,0}(t) &= \rho_{\pm,0}(0) \exp \left(-i n \alpha \epsilon_\Delta^2 t - \Gamma(n,\xi) t \right)
\end{align}
This shows that indeed the chosen entangled state yields eq.~\ref{off-diagonal time evolution}. We just need to calculate the expression for $\Gamma(n,\xi)$, which we do next.

We calculate the sum over each of the three terms in eq.~\ref{master equation supp mat} for a fixed distance of ions $|j-k|=x$ and then regard the coefficients of $C(0,x)$. There are $(n-x)$ pairs of ions with a distance $x$ between them, which can be seen by moving a fixed distance along figure \ref{fig chain of n ions 2}. For $x>0$ each pair is counted twice because $j$ will be the right one and the left one once. We distinguish three cases:

\subsection*{case 1) $x$ is odd}
In each pair there is one spin in state $\epsilon_+$ and one in state $\epsilon_-$.
\begin{align}
\sum_{|j-k|=x} J_z^{(j)} \rho_{\pm,0} J_z^{(k)} &= 2 (n-x) \epsilon_0^2 \rho_{\pm,0}\\
\sum_{|j-k|=x} J_z^{(j)} J_z^{(k)} \rho_{\pm,0} &= 2 (n-x) (\epsilon_0^2 - \epsilon_\Delta^2) \rho_{\pm,0}\\
\sum_{|j-k|=x} \rho_{\pm,0} J_z^{(j)} J_z^{(k)} &= 2 (n-x) \epsilon_0^2 \rho_{\pm,0}
\end{align}
For the last equation, note that $j$ is each ion in the pair once and therefore give once $\epsilon_+$ and once $\epsilon_-$. Furthermore $\epsilon_+ + \epsilon_- = 2 \epsilon_0$ for each pair.
\begin{align}
\frac{1}{2} \sum_{|j-k|=x} C(0,|j-k|) \left( 2J_z^{(j)} \rho_{\pm,0} J_z^{(k)} - J_z^{(j)} J_z^{(k)} \rho_{\pm,0} - \rho_{\pm,0} J_z^{(j)} J_z^{(k)} \right) = +(n-x) \epsilon_\Delta^2\; C(0,x) \rho_{\pm,0}
\end{align}
Note that the contribution of pairs which are in opposite states is \emph{positive}, i.e.~reduces the dephasing rate.

\subsection*{case 2) $x$ is even}
In each pair the ions are in the same state. There is an equal number of  $(\epsilon_+, \epsilon_+)$ pairs and $(\epsilon_-, \epsilon_-)$ pairs because (moving along the chain in figure \ref{fig chain of n ions 2}) the pairs start with one type and finish with the other ($n$ is even).
\begin{align}
\sum_{|j-k|=x} J_z^{(j)} \rho_{\pm,0} J_z^{(k)} &= 2(n-x) \epsilon_0^2 \rho_{\pm,0}\\
\sum_{|j-k|=x} J_z^{(j)} J_z^{(k)} \rho_{\pm,0} &= 2(n-x)(\epsilon_0^2+\epsilon_\Delta^2) \rho_{\pm,0}\\
\sum_{|j-k|=x} \rho_{\pm,0} J_z^{(j)} J_z^{(k)} &= 2(n-x)\epsilon_0^2 \rho_{\pm,0}
\end{align}
\begin{align}
\frac{1}{2} \sum_{|j-k|=x} C(0,|j-k|) \left( 2J_z^{(j)} \rho_{\pm,0} J_z^{(k)} - J_z^{(j)} J_z^{(k)} \rho_{\pm,0} - \rho_{\pm,0} J_z^{(j)} J_z^{(k)} \right) = -(n-x) \epsilon_\Delta^2\; C(0,x) \rho_{\pm,0}
\end{align}
Note that the contribution of pairs which are in the same state is \emph{negative}, i.e.~increases the dephasing rate.

\subsection*{case 3) $x=0$}
For $x=0 \Leftrightarrow j=k$ there are n summands (which should not have the factor of 2 from the other cases).
\begin{align}
\sum_{|j-k|=x} J_z^{(j)} \rho_{\pm,0} J_z^{(k)} &= n \epsilon_0^2 \rho_{\pm,0}\\
\sum_{|j-k|=x} J_z^{(j)} J_z^{(k)} \rho_{\pm,0} &=  n(\epsilon_0^2+\epsilon_\Delta^2) \rho_{\pm,0} \\
\sum_{|j-k|=x} \rho_{\pm,0} J_z^{(j)} J_z^{(k)} &= n \epsilon_0^2 \rho_{\pm,0}
\end{align}
\begin{align}
\frac{1}{2} \sum_{|j-k|=x} C(0,|j-k|)\left( 2J_z^{(j)} \rho_{\pm,0} J_z^{(k)} - J_z^{(j)} J_z^{(k)} \rho_{\pm,0} - \rho_{\pm,0} J_z^{(j)} J_z^{(k)} \right) = -\frac{n}{2}\epsilon_\Delta^2 \rho_{\pm,0}
\end{align}

We have now calculated all coefficients for the spatial correlation functions $C(0,x)$ in the decoherent part of the master equation and can now write it down with an analytical expression for the dephasing rate $\Gamma(n,\xi)$:
\begin{align}
\dot \rho_{\pm,0} &= -i n \alpha \epsilon_\Delta^2 \rho_{\pm,0} - \Gamma(n,\xi) \rho_{\pm,0}\\
\rho_{\pm,0}(t) &= \rho_{\pm,0}(0) \exp \left(-i n \alpha \epsilon_\Delta^2 t - \Gamma(n,\xi) t \right)
\end{align}
The dephasing rate $\Gamma(n,\xi)$ is given by the contributions for all $x$, which we divide into one summand for $x=0$, the sum over odd $x=2x_c-1$ and the sum over even $x=2x_c$:
\begin{align}
\Gamma(n,\xi)&=\left( \frac{n}{2} - \sum_{x_c=1}^{n/2} (n-2x_c+1) C(0,2x_c-1) + \sum_{x_c=1}^{n/2} (n-2x_c) C(0,2x_c) \right) \epsilon_\Delta^2 \label{quadrupole measurement dephasing rate of x and xi 2}
\end{align}
We have now derived eq.~\ref{dephasing rate for general C}. Inserting an exponential correlation function $C(0,|j-k|)=\exp(-|j-k|/\xi)$ yields the analytical expression of eq.~\ref{dephasing rate of n and xi}:
\begin{align}
\Gamma(n,\xi)=\frac{2 e^{\frac{1}{\xi }}-2 e^{\frac{1-n}{\xi }}-n+e^{2/\xi } n}{2 \left(1+e^{\frac{1}{\xi }}\right)^2} \epsilon_\Delta^2
\end{align}
This is the dephasing rate for exponentially decaying partial correlations. For uncorrelated decoherence $\xi \rightarrow 0$ all collective terms $j\neq k$ vanish and only case 3 gives a non-vanishing term. The uncorrelated dephasing rate is therefore
\begin{align}
\Gamma_{\rm uc}=n \epsilon_\Delta^2/2
\end{align}

\section{Oscillating spatial correlations facilitate a decoherence-free GHZ state}
\label{app oscillating correlations}
\begin{figure}[h!]
\includegraphics[scale=1]{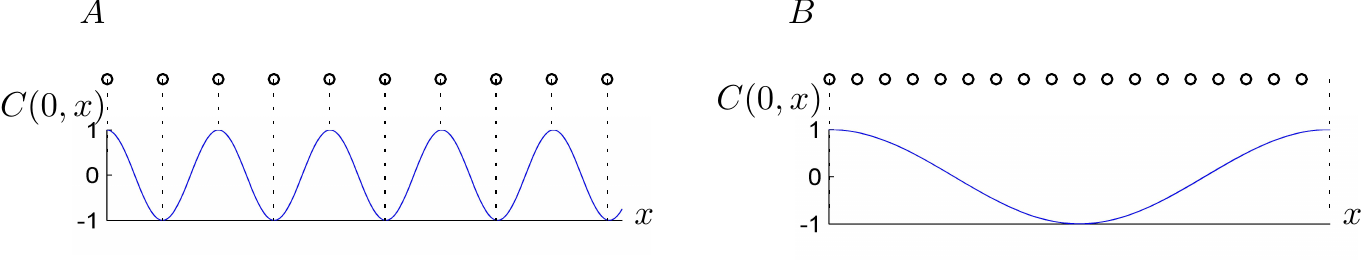}
\caption{Two arrangements of ions in oscillating spatial correlations. Both arrangements achieve Heisenberg scaling of the GHZ state: A) For small oscillation length the ions are arranged to meet half the oscillation length. B) For long oscillation length the whole array is arranged to match up one oscillation length.}
\label{fig coscorrelations}
\end{figure}

Up to here we have considered perfect spatial correlations $C(0,x)=1$ and decaying spatial correlations $C(0,x)=\exp(-x/\xi)$. Both are positive functions for all $x$ and the $n^2$ scaling of the dephasing rates found experimentally\cite{Monz2011} for GHZ states indicates that in ion traps these two functional forms are good approximations for the noise correlations. However, it is also physically possible for the spatial correlations to take the homogeneous form $C(0,x)=\cos(k_s x)$, where points of certain distances have noise with negative correlations. For example in reference \cite{Jeske2013formalism} an environmental model is discussed that yields cosine shaped spatial correlations. In such an environment GHZ states can be engineered to be within a decoherence-free subspace in two ways assuming the spatial oscillation length is known. One way is to place the ions at half the oscillation length of the environmental spatial correlations (figure \ref{fig coscorrelations}A); the other way is to match the array length $L$ with the oscillation length $L=2 \pi/k_s$ (figure \ref{fig coscorrelations}B).

For the GHZ state's relevant coherence $r_3=\ket{111...}\bra{000...}$ all operator pairings $j,k$ in the master equation have the same effect. The only difference comes from the cosine shaped correlation function:
\begin{align}
\dot r_3 &=  i [r_3, H_s] + \sum_{jk} \cos(k_s d |j-k|) \left( -\sigma_z^{(j)} \sigma_z^{(k)} r_3 - r_3 \sigma_z^{(j)} \sigma_z^{(k)} + \sigma_z^{(j)} r_3 \sigma_z^{(k)} + \sigma_z^{(k)} r_3 \sigma_z^{(j)} \right) \label{r_3 master eq for cos correlation}
\end{align}
where $d$ is the distance between ions.

In the arrangement of figure \ref{fig coscorrelations}A the correlation function becomes effectively $\cos(k_s d |j-k|)=(-1)^{|j-k|}$, i.e. alternates the sign with increasing distance. This recreates the effect of the coherence
$$ \rho_{\pm,0}=\ket{m_-,m_+,m_-,m_+,...}\bra{m_1,m_1,m_1,...,m_1}$$
of the previous section, where the alternating sign comes from the arrangement of the ions and the correlation function is always the same sign (positive). Here the alternating sign comes from the correlation function and the operators produce the same term for all pairs of $j$ and $k$. From this equivalence one finds that the arrangement of figure \ref{fig coscorrelations}A leads to Heisenberg scaling for the GHZ state, even if the cosine correlation function has an additional exponentially decaying envelope.

In the arrangement of figure \ref{fig coscorrelations}B the length of the array matches the spatial oscillation length of the environmental correlations. Regarding eq.~\ref{r_3 master eq for cos correlation} we group the pairs of the first ion $j=1$ with all the other ions (including the autocorrelation $j=k=1$). We find that for each positive contribution $C(0,(1-k)d)>0$ there is an equal negative contribution from the $k$-value $n/2$ further down the chain. The sum of all contributions of $j=1$ therefore cancels (assuming an even number $n$ of ions). The same argument applies for the sum of all contributions for $j=2$ or any other value of $j$. For large numbers $n$ of ions these summations for one value of $j$ approach an integration over one oscillation length of a cosine, which illustrates the summation to zero even better.

In both arrangements of figure \ref{fig coscorrelations} the GHZ state $(\ket{111...} +\ket{000...})/\sqrt{2}$ turns out to be a decoherence-free state or dark state for all $n$ and frequency measurements with it will therefore show Heisenberg scaling. The difficulty of an experimental implementation of this is to find and map out a noise environment with cosine spatial correlations.

After discussing purely positive and oscillating spatial correlations we would like to point out that purely negative correlation functions, i.e.~perfect anti-correlations are impossible due to the necessity of positive autocorrelations $j=k$ and multipartite correlation rules. Particularly negative noise correlations between position a and b combined with negative correlations between position b and c require positive correlations between position a and c.

\section*{References}
\bibliographystyle{ieeetr}
\bibliography{publication}

\end{document}